\documentstyle[aps,preprint,tighten,epsfig]{revtex}
\def\beq{\begin{equation}}
\def\enq{\end{equation}}
\def\beqa{\begin{eqnarray}}
\def\enqa{\end{eqnarray}}
\def\MeV{\nobreak\,\mbox{MeV}}
\def\GeV{\nobreak\,\mbox{GeV}}

\def\ka{\kappa}
\def\la{\lambda}
\def\ga{\gamma}
\def\Ga{\Gamma}

\def\rh{\rho}
\def\si{\sigma}

\def\de{\delta}
\def\al{\alpha}

\def\lb{\label}
\def\nn{\nonumber}
\newcommand{\rag}{\rangle}
\newcommand{\lag}{\langle}

\newcommand{\rf}{\ref}
\newcommand{\ct}{\cite}
\begin{document}
\title{\sc  Semileptonic $D$ decay into scalar mesons: a QCD sum rule approach}
\author { H.G. Dosch$^1$, E.M. Ferreira$^2$,  
F.S. Navarra$^3$ and M. Nielsen$^3$}
\address{  $^1$Institut f\"ur Theoretische Physik, Universit\"at Heidelberg\\
 Philosophenweg 16, D-6900 Heidelberg, Germany\\[0.1cm]
 $^2$Instituto de F\'{\i}sica, Universidade Federal do Rio de  
Janeiro \\
C.P. 68528, Rio de Janeiro 21945-970, RJ, Brazil\\
 [0.1cm]
 $^3$Instituto de F\'{\i}sica, Universidade de S\~{a}o Paulo\\
  C.P. 66318,  05315-970 S\~{a}o Paulo, SP, Brazil}
\maketitle
\vspace{1cm}
\begin{abstract}

Semileptonic decays of $D$-mesons into scalar hadronic states are 
investigated. Two extreme cases are considered: a) the meson decays 
directly into an uncorrelated scalar state of two two mesons  and 
b) the decay proceeds via resonance formation. QCD sum rules including 
instanton contributions are used to calculate total and differential 
decay rates under the two assumptions. 
\end{abstract} 

\bigskip

\bigskip

PACS Numbers~ :~ 12.38.Lg , 13.30.Ce ,
14.65.Dw\newpage

\section{Introduction}
Low lying scalar mesons are an  old problem in hadron physics, see the
review by Spanier and T\"ornqvist on scalar mesons in \ct{PDG00} and the
literature quoted there. In a recent analysis \cite{Ait01,e791} of $D$- and
$D_s$-meson nonleptonic decays  distinct signals for strong enhancements in 
$S$-wave $\pi \pi$ and $K \pi$ channels have been observed,  reviving the
interest for these states.  In the following we shall refer to the signal in
the $\pi \pi$ channel as $\si$ and in the $K \pi$ chanel as $\ka$. The
enhancements can be well described by a Breit-Wigner-type resonance form,  with
the  correct threshold behaviour
\beq
\rh_{X,BW}(s) = \frac{\Ga_X(s) m_X}{(s-m_X^2)^2+m_X^2\Ga_X(s)^2}\;,
\lb{BW}
\enq
where the subscript $X$ stands for $\si$ or $\ka$ . 
The correct threshold behaviour is guaranteed through the $s$-dependent
width
\beq
\Ga_X(s) = \Ga_{0X} \frac{\la^{1/2}(s,m_a^2,m_b^2)}
{\la^{1/2}(m_X^2,m_a^2,m_b^2)}{m_X^2\over s}.
\lb{Ga}
\enq
Here $m_a$ and $m_b$ are the masses of the mesons in the decay channel and
$\la(x,y,z)=x^2+y^2+z^2-2 x y - 2 x z - 2 y z$.
In \cite{e791} the following parameters were found:\\
1) for the $S$-wave $\pi\pi$ channel, $m_a=m_b=m_\pi$,
\beq
m_\si=  0.478 \pm 0.024 \pm 0.017 \GeV;
 ~~~ \Ga_{0\si} = 0.324\pm0.042\pm0.021 \GeV.
\lb{ressi}
\enq
2) for the $S$-wave $K\pi$ channel, $m_a=m_K,~m_b=m_\pi$, 
\beq
m_\ka= 0.797 \pm 0.019 \pm 0.042 \GeV;
 ~~~ \Ga_{0\ka} = 0.410 \pm 0.043 \pm 0.085 \GeV.
\lb{reska}
\enq

A theoretical analysis \ct{GNPT00} of the non-leptonic decay $D \to \sigma
\pi$ is in agrement with the experimental data. The semileptonic
$D$-decays offer, in principle, much cleaner samples than the nonleptonic
decays since there occur no problems connected with the presence of a third
strongly interacting particle. In the non-strange channel important 
experimental information comes indeed from the analysis of the final state
interaction of the pions from leptonic $K$-decays. For the case of strange
mesons, the analysis of $D$ decays can play a similarly important role. In a
recent paper by the FOCUS collaboration \ct{Lin02} clear evidence was found
that in the semileptonic  decay $D^+\to K^- \pi^+\mu^+\nu$ the $D$-meson does
not decay exclusively into the hadronic vector channel, but that there is
interference with a scalar contribution. In this paper we estimate the decay
rates of semileptonic $D$-decays into the scalar $K \pi$ and $\pi \pi$
channels. We use the method of QCD  sum rules \cite{SVZ79} which has been
successfully applied to several semileptonic decay processes. For a recent
review see \cite{CK00} and the literature quoted there. 

Even if (broad) resonances in the hadronic decay channels exist, it is not
clear whether they are due to interactions on the quark level, or 
if they are rather an effect of interactions in the purely hadronic channel,
see for instance \ct{JOP00}. Here a theoretical
analysis can be very helpful. The semileptonic decay is supposed  to occur on
the quark level and therefore the decay rate should depend  crucially on the
direct coupling of the resonances to the quark currents.   We 
consider two limiting cases 
\begin{figure} \setlength{\unitlength}{0.5cm}
\begin{center} \begin{picture}(8,5) \put(2,0){\line(1,1){2}}
\put(4,2){\line(1,-1){2}} \put(2,0){\line(1,0){4}} \put(8,0){\line(2,1){2}}
\put(8,0){\line(2,-1){2}} \thicklines
\put(0,0){\line(1,0){2}}
\put(4,2){\line(0,1){2}}
\put(6,0){\line(1,0){2}}
\put(1,0.2){$D$}
\put(7,0.2){$X$}
\put(4.2,3){$j_\mu$}
\put(4,0.2){$q_1$}
\put(9,0.8){$a$}
\put(9,-0.4){$b$}
\put(2.5,1){$c$}
\put(5.2,1){$q_2$}
\end{picture}
\end{center}
\vspace{0.5cm}
\caption{Schematic picture  of the semileptonic vertex of a $D$-meson decay.
$c,~q_1,~q_2$ are quark lines, $j_\mu$ denotes the weak current. $X$ might be
either  a (broad) resonance or represent an uncorrelated two-meson state of
mesons $a$ and $b$.} \lb{graph} 
\end{figure}
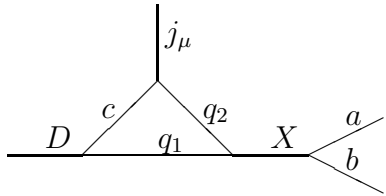

\begin{itemize} \item the observed two-meson final
state couples to the corresponding quark-antiquark state only through the
$\sigma$ or $\kappa$ resonance. 
\item 
the quark antiquark state couples to uncorrelated final mesons in an $S$-wave.
\end{itemize} 
Since the signals we investigate in this  paper are very broad the finite 
width has to be taken into account.
Otherwise our analysis is based on the same principles and assumptions as the
used in the sum-rule analysis of other semileptonic $D$-decays~\cite{BBD91}. 
We therefore refer to this paper for  details and 
clarifications.

\section{Kinematics}

We investigate the semileptonic decays
\beq
D\to X \ell \bar \nu_\ell\;,
\enq
where $X$ might be the $\si,~ \ka$ or an 
uncorrelated $\pi \pi$ or $K\pi$
pair in the $S$-state, as depicted in Fig.~\rf{graph}.

The semileptonic decay of a $D$-meson with momentum $p_D$ into a scalar state
with total momentum $p_X$ and invariant mass $\sqrt{s}=\sqrt{p_X^2~}$ is
described by the two form factors of the matrix element of the weak current
\beq
\lag X |j_\mu| D\rag =(p_D+p_X)_\mu f_{+X}(t) +(p_D-p_X)_\mu f_{-X}(t)\;,
\lb{ff}
\enq
where $t=(p_D-p_X)^2$.
In the decay rate the form factor $f_{-X}$ is multiplied by the difference of 
the lepton
masses and hence is negligible for both $e$ and $\mu$ decays.

The differential semileptonic decay rate is given by
\beq
\frac{d^2\Ga(s,t)}{ds~dt} =
\frac{G_F^2|V_{cq_2}|^2}{192\pi^3m^3_D} \la^{3/2}(m_D^2,s,t)f_{+X}^2(t) 
\frac{\rh_X(s)}{\pi}\;,
\lb{gadif}
\enq
where $G_F$ is the Fermi coupling constant and $V_{cq_2}$ the CKM transition
element from the charmed quark to the quark $q_2$.
The total width is
\beq
\Ga = \int_{(m_a+m_b)^2}^{m_D^2}ds\int_0^{(m_D-\sqrt{s})^2} dt~\frac{
d^2\Ga(s,t)}{ds~dt}\;.
\lb{gatot}
\enq

The spectral distribution in the invariant mass $\sqrt{s}$ of the hadronic
final state is given by
\beq 
\frac{d \Ga}{d\sqrt{s}} =
2 \sqrt{s~}\int_0^{(m_D-\sqrt{s})^2} dt~\frac{ 
d^2\Ga(s,t)}{ds~dt} . \lb{gaspec} 
\enq 

\section{Sum rules}

The $D$ meson in the initial state is interpolated by the pseudoscalar
current
\beq
j_D(x) = \bar{c}(x)i\ga_5 q_1(x)\;,
\lb{intD}
\enq
where $c$ is the field of the charmed quark and 
$q_1$ that of an up or down quark, summation over spinor and colour indices
being understood but not indicated explicitly. The final hadronic  state $X$
is interpolated by the scalar current 
\beq
j_X(x) = \bar{q}_1(x) q_2(x)\;,
\lb{intscal}
\enq
where $q_2$ represents a light quark field for $X=\si$, and  a strange
quark field for $X=\ka$.
The semileptonic decay rate is obtained from the time ordered product
of the two interpolating fields in Eqs.~(\rf{intD}) and (\rf{intscal}) and the
weak current $j^W_\mu=\bar{q}_2 \gamma_\mu(1-\gamma_5) c$
\beq
T_{\mu X}(p_D^2,p_X^2,t) = i^2\int d^4x d^4y\, \lag0|{\rm T}[ j_D(x) j^W_\mu(0)
j_X(y)]|0\rag e^{i(p_D.x-p_X.y)}\;.
\lb{3point}
\enq

In order to select the semileptonic decay rates into the lowest lying
hadrons we insert intermediate states and obtain the
following double dispersion relation in the phenomenological side 
\beqa
T_{\mu X}^{phen} (p_D^2,p_X^2,t)&=&\frac{1}{\pi^2} \int ds_D\,ds_X\,
\lag 0 | j_D(0)|D\rag
\lag D|j^W_\mu(0)|X\rag \lag X|j_X(0)|0\rag \nn\\
&&\times \frac{1}{s_D-p_D^2}
\frac{1}{s_X^2-m_X^2}
\mbox{ + contributions of higher resonances}\;.
\lb{dpr}
\enqa
Introducing
\beqa
\lag 0 | j_D(0)|D\rag &=& A_D \rh_D(s_D);~~~~p_{D}^2=s_D\;,\lb{constd}\\
\lag X|j_X(0)|0\rag &=&A_X \rh_X(s_X);~~~~p_{X}^2=s_X\;,
\lb{const}
\enqa
and using Eq.~(\rf{ff}) we obtain
\beqa
T_{\mu X}^{phen}(p_D^2,p_X^2,t)&=&\frac{1}{\pi^2} \int ds_D\,ds_X\, A_D A_X
\frac{\rh_D(s_D)}{s_D-p_D^2}~\frac{\rh_X(s_X)}{s_X-p_X^2}\nn \\
&&\times \Big(f_{+X}(t,s_D,s_X)(p_D+p_X)_\mu+f_{-X}(t,s_D,s_X)
(p_D-p_X)_\mu\Big)\nn\\
&&+\mbox{ contributions of higher resonances}\;.
\lb{3point2}
\enqa

In the following we concentrate on the relevant form factor
$f_+$ and introduce the factor $T_+$ which multiplies the vector
$(p_D+p_X)_\mu$ 
\beqa
T_{+X}^{phen}(p_D^2,p_X^2,t)&=&\frac{1}{\pi^2} \int ds_D\,ds_X\,
A_D A_X \frac{\rh_D(s_D)}{s_D-p_D^2}~\frac{\rh_X(s_X)}{s_X-p_X^2}~
f_{+X}(t,s_D,s_X) \nn\\
&&+\mbox{ contributions of higher resonances}\;.
\lb{3point3}
\enqa
For the  $D$-meson the density $\rho_D(s_D)$ introduced in Eq.~(\rf{constd}) is
given by
\beq
\rh_D(s_D)= \pi \de(s_D-m_D^2)\;,
\enq
and we obtain 
\beq A_D=f_D m_D^2/m_c  \; , 
\lb{Af}
\enq
 $f_D$ being the $D$-meson decay constant in the conventional notation.

For the density $\rh_X$ we use the two extreme
ans\"atze mentioned above:
\begin{itemize}
\item The quark-antiquark current $j_X$ in the $J=0^+$ state couples to the
$\sigma$ or $\kappa$-resonance described by the Breit-Wigner distribution
(\rf{BW})  
  \beq 
\rh_{X}(s)=\rh_{X,BW}(s)\;. \lb{rho1} 
\enq 
\item
The quark-antiquark current $j_X$ couples to an uncorrelated  meson pair,
the density being described by the density of two-particle phase space  
\beq 
\rh_X(s) = \frac{\pi}{16 \pi^2} \frac{\la(s,m_a^2,m_b^2)^{1/2}}{s} ,
 \lb{rho2} 
\enq 
where $m_a$, $m_b$ are the masses of the mesons in the final state. 
\end{itemize}

The three-point function can be
evaluated by perturbative QCD if the external momenta are in the deep Euclidean
region
\beq
p_D^2\ll (m_c+m_1)^2,~~~ p_X^2 \ll (m_1+m_2)^2, ~~~ t \ll (m_c+m_2)^2\;.
\lb{cond} 
\enq
In order to approach the not-so-deep-Euclidean region and to 
get more information on the nearest physical singularities, 
nonperturbative power corrections are added to the perturbative contribution
\beq
T^{theor}_{+X}(p_D^2,p_X^2,t)= {1\over\pi^2}\int_{s_{D,th}}ds_D
\int_{s_{X,th}} ds_X\,
\frac{\si_{+X}(s_D,s_X,t)}{(s_D-p_D^2)(s_X-p_X^2)}
+ \sum_{ij}\frac{C_{ij}}{(p_D^2)^i (p_X^2)^j}\lag O_{ij}\rag\;.
\lb{power}
\enq
The perturbative contribution is contained in the double spectral function
$\si_{+X}$. The Wilson coefficients $C_{ij}$ multiplying the power
corrections can be evaluated in perturbative QCD. The operators 
$O_{ij}$ occur in the operator expansion of the time ordered product  
Eq.~(\rf{3point}); their vacuum expectation values, the condensates   $\lag
O_{ij}\rag$, are introduced as phenomenological parameters.

In order to suppress the condensates of higher dimension and at the same time
reduce the influence of higher resonances, the series in Eq.~(\rf{power}) is
Borel improved,  leading to the  mapping 
\beq
f(p^2) \to \hat f(M^2),~~~~~\frac{1}{(p^2-m^2)^n} \to \frac{(-1)^n}{(n-1)!}
\frac{e^{-m^2/M^2}}{(M^2)^n}\;.
\enq
Furthermore, we make the usual assumption that the contributions of  higher
resonances are well approximated by the perturbative expression
\beq
\frac{1}{\pi^2}\int_{s_{0D}}^\infty ds_D\int_{s_{0X}}^\infty ds_X\,
\frac{\si_{+X}(s_D,s_X,t)}{(s_D-p_D^2)(s_X-p_X^2)}\; ,
\enq 
with appropriate continuum thresholds $s_{0D}$ and $s_{0X}$.
By equating the Borel transforms of the phenomenological expression
in Eq.(\rf{3point3}) and that of the ``theoretical expression'', Eq.~(\rf{power}),
we obtain the sum rule
\beqa
\lefteqn{f_{+X}(t,m_D^2,\bar s_X)
f_D {m_D^2\over m_c}e^{-m_D^2/M_D^2}{A_X\over\pi}\, \int_{(m_a+m_b)^2}^{s_{0X}}
ds_X\,
 \rh_{X}(s_X) e^{-s_X/M_X^2}} \nn \\
&&=\frac{1}{\pi^2} \int_{s_{D,th}}^{s_{0D}} ds_D\int_{s_{X,th}}^{s_{0X}}
ds_X\,
\si_{+X}(s_D,s_X,t)e^{-s_D^2/M_D^2}\,e^{-s_X/M_X^2}\nn  \\
&& ~~~~~~~ + \hat K_{+X}(M_D^2,M_X^2,t)~ + \hat K_{+I}(M_D^2,M_X^2,t)\;,
\lb{power2} 
\enqa
where $\bar s_X$ is some value below $s_{0X}$. In the zero 
width approximation we have of course $\bar s_X=m_X^2$. $\hat K_{+X}$ are the
Borel transforms of the nonperturbative expressions due to the condensates and 
$\hat K_I $ is an approximation to the instanton contributions, which might be
important in the scalar channel \cite{shu}.

The decay constant $f_D$ and the coupling
$A_X$ defined in Eqs.~(\rf{constd}), (\rf{const}), and(\rf{Af})  can also be
determined  by sum rules
obtained from the appropriate two-point functions. Using the same procedure as
described above we arrive at \beq
{m_D^4\over m_c^2}\Big(f^{theor}_D(M^2)\Big)^2 e^{-m_D^2/M^2}=
\int_{m_c^2}^{s_{0D}}  ds\,\si_D(s) e^{-s/M^2}
+ \hat K_D(M^2)\;,\lb{twopointd}
\enq
and
\beqa
\lefteqn{\Big(A^{theor}_X (M^2)\Big)^2
\Big({1\over\pi}\int_{(m_a+m_b)^2}^{s_{0X}} ds \rh_X(s)   e^{-s/M^2}\Big)=}\nn
\\ &&~~~~ 
\int_{m_{q_2}^2}^{s_{0X}} ds\, \si_X(s)e^{-s/M^2} +\hat K_X(M^2)+
\hat K_I(M^2)\;.
\lb{twopoint}
\enqa

The analysis of the two-point function for the scalar mesons, and the
explicit expressions for the functions occuring in Eqs.~(\rf{power2}),
(\rf{twopointd}) and (\rf{twopoint}) are given in  Appendices A and B 
respectively.

The final sum rule for the form factor is obtained from Eq.~(\rf{power2}) 
by inserting for $f_D$ and $A_X$ the expressions   
$f^{theor}_D$ and $A^{theor}_X$ of Eqs.~(\rf{twopointd}) and (\rf{twopoint}) 
\beqa
\lefteqn{f_{+X}(t,m_D^2,\bar s_X) =}\nn \\
&& e^{m_D^2/M_D^2} 
\left( {m_D^2\over m_c}f_D({M_D'}^2) A_X({M_X'}^2){1\over\pi}
 \int_{(m_a+m_b)^2}^{s_{0X}}ds_X\,\rh_X(s_X) e^{-s_X/M_X^2}\right)^{-1}\nn \\
&\times&\bigg(\int_{s_{D,th}}^{s_{0D}} ds_D\int_{s_{X,th}}^{s_{0X}}
ds_X\,
\si_{+X}(s_D,s_X,t)e^{-s_D^2/M_D^2}\,e^{-s_X/M_X^2}\nn \\
&+& \hat K_{+X}(M_D^2,M_X^2,t) \;+\; \hat K_{+I}(M_D^2,M_X^2,t)\bigg)\;.
\lb{sr3}
\enqa

The radiative corrections for the scalar and pseudoscalar channels are 
known to be large \ct{bro}.
They are expected to be large in the three-point function too. By inserting
the sum rule expressions for the two-point functions, Eqs.~(\rf{twopointd})
and (\rf{twopoint}), in the denominator
of the sum rule for the three-point function, Eq.~(\rf{sr3}), we expect, at 
least, a partial cancellation of these corrections \cite{BBB98,BBFG95}.

\section{Evaluation of the sum rules and results}

The sum rule Eq.~(\rf{sr3}) is evaluated in the same way as described 
in \ct{BBD91}, and we only sketch the main steps of this 
evaluation. In the complete theory,  the
right hand side of Eq.~(\rf{sr3}) should not depend on the Borel
variables $M^2$. However, in  a truncated treatment there will always  be some
dependence left.  Therefore, one has to work in a region 
where the approximations made are supposedly acceptable and where 
the result depends only moderately on the Borel variables . To decrease the 
dependence of the results on the Borel variables $M^2$ , we take
them in the two-point functions at half the
value of the corresponding variables in the three-point sum rules, 
{\it i.e.}, in Eq.(\rf{sr3}) we put
\beq {M_D'}^2= M_D^2/2 ~~~~ {\rm and} ~~~~ {M_X'}^2=  M_X^2/2\;.
\lb{m23}
\enq
We furthermore choose
\beq
\frac{M_X^2}{M_D^2} = \frac{m_X^2}{m_D^2-m_c^2}\;. \lb{mm}
\enq
We have checked that the results do not depend crucially on this particular
choice.  If the momentum transfer $t$ to the lepton pair is larger than a
critical  value $t_{cr}$,
non-Landau singularities have to be taken into account \ct{BBD91}. 
Since anyhow we have
to stay away from the physical region , {\it i.e.} we must have
$t \ll (m_c+m_1)^2$, we limit our calculation to the region $0 < t < t_{cr}$.
In this  range the t-dependence can be obtained from the sum rule (\rf{sr3})
directly. It can be fitted by a monopole, 
and extrapolated to the full kinematical region.

Since we do not take into account radiative corrections we choose the QCD
parameters at a fixed renormalisaton scale of about $1~\GeV^2$:
the strange and
charm mass $m_s=0.16\,\GeV$, $m_c=1.3\,\GeV$, the up and down quark masses are
put to zero.
We take for the non-strange quark condensate
$\langle\overline{q}q\rangle\,=\,-(0.24)^3\,\GeV^3$ , 
for the strange
quark
condensate $\langle\overline{s}s\rangle\,=\,0.8\,\langle\overline{q}q\rangle$,
and for the  mixed quark-gluon condensate $\langle\overline{q}\sigma G
q\rangle\langle\overline{q}q\rangle=m_0^2 \langle\overline{q}q\rangle$ with 
$m_0^2=0.8\,\GeV^2$.

For the continuum threshold in the $D$-channel we take from \cite{BBD91}
$s_{0D}=6~\GeV^2$. The standard value in the $X$ channel
would be $s_{0X} \approx (m_X+0.5~\GeV)^2$, yielding $s_{0\sigma}\approx 1
\GeV^2$ and  $s_{0\kappa}\approx 1.6  \GeV^2$. As an additional condition we
use the mass constraint from the two-point function, as described in
Appendix A.

We start with the decay $D \to \ka \ell \bar \nu_\ell$ 
and first consider the case where the scalar quark current (\rf{intscal}) 
couples directly to the $\kappa$-signal through a Breit-Wigner distribution
(\ref{rho1}).  In Fig.~2 we show the different  contributions to the form
factor $f_{+\ka}(0) $ in Eq.~(\rf{sr3}), as a function of the Borel variable
$M_D^2$, using  the continuum threshold $s_{0\kappa}= 1.6 \GeV^2$. As the lower
limit for the fiducial region in $M_D^2$ we take that value of $M_D^2$ where
the perturbative contribution is one half of the total contribution.
As upper limit we take  the value $M_D^2=15 \GeV^2$, which is motivated in
Appendix A. For such a high value of the Borel parameter the result is very
stable but it is largely determined by the choice of the continuum model. The
instanton contribution in the fiducial region is completely negligible; the 
five dimensional mixed condensate is  strongly suppressed compared to
the three dimensional quark condensate. In the Borel variable of the
$X$-channel the fiducial region corresponds approximately to the range
$1.7 \GeV^2 \leq M_X^2 \leq 5 \GeV^2$.
\begin{figure} \label{fig2}
\begin{center}
\epsfysize=8.0cm
\epsffile{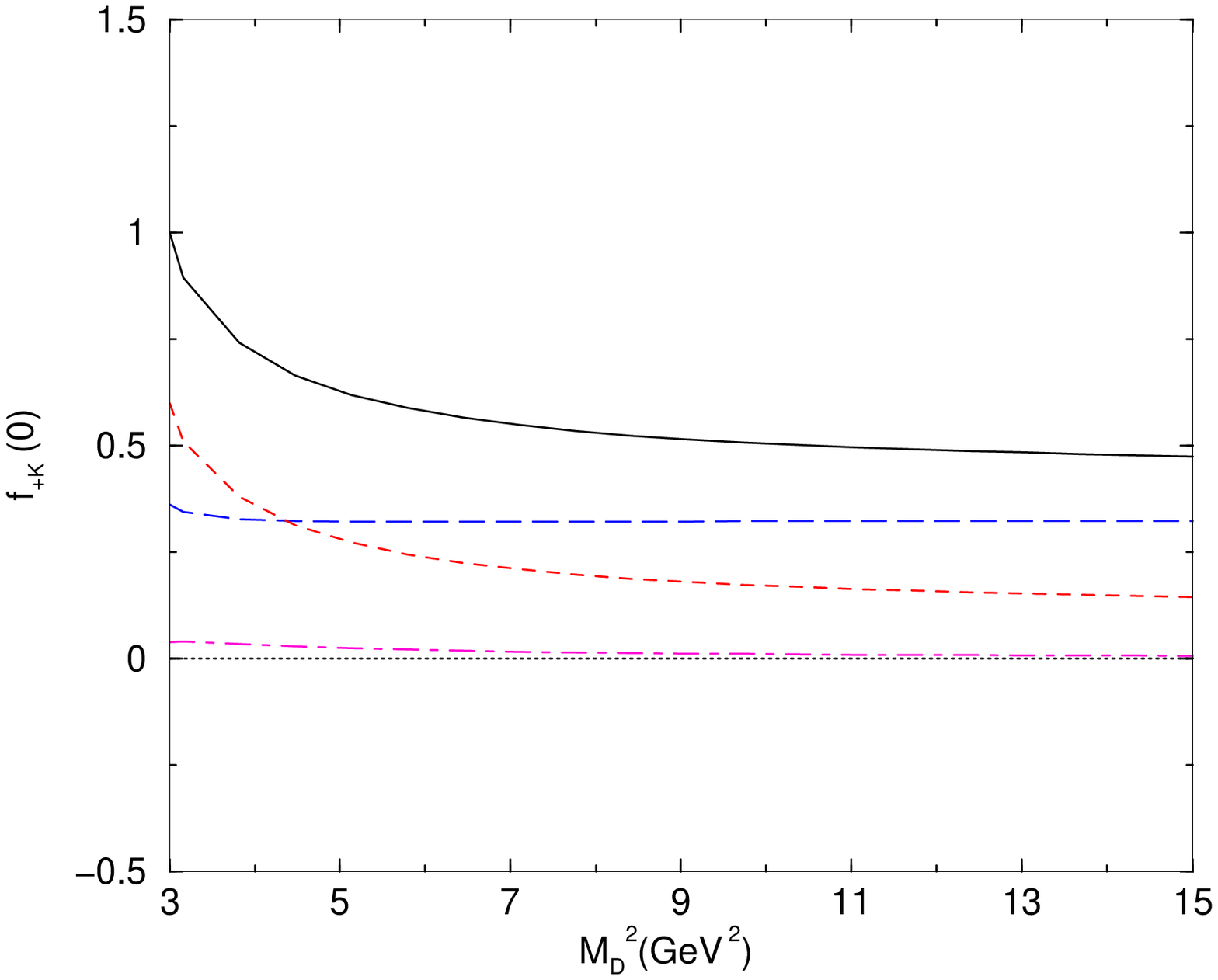}
\caption{Dependence of the form factor $f_{+\ka}$ at
$t=0$ on the Borel variable $M_D^2$. Here the decay is assumed to proceed
through resonance formation and the state density is given by Eq.\protect
(\rf{rho1}). Solid curve: total contribution; long-dashed: perturbative; 
dashed:
quark condensate; dot-dashed mixed condensate; dotted : instanton
contribution.}  \end{center} 
\end{figure}

In the range  $0\leq t \leq 0.5 \; \GeV^2 $ no non-Landau singularities
occur for our choices of the continuum thresholds. The  momentum
dependence of $f_{+\ka}$ 
can, in this $t$-range, be very well  approximated by a monopole expression
\beq
f_{+\ka}(t)={f_{+\ka}(0)\over 1-{t\over M_P^2}}\;,
\lb{mo}
\enq
and extrapolated to the full physical region.

For $1.4~\GeV^2 \leq s_{0\kappa} \leq 1.8~\GeV^2$ and for values of
$M_D^2$ discussed above, we find for the form factor at $t=0$ 
\beq 
0.48\leq f_{+\ka}(0) \leq 0.55\,,
\enq 
and for the pole mass,
\beq
1.9~\GeV\leq M_P \leq 2.2~\GeV\,.
\enq
The pole mass is considerably smaller  than the mass of the charmed 
pseudovector-meson $D_{s1}(2536)$ which would fit into the $t$-channel. In the
$D\to K \ell \nu$ decay the pole mass also came out to be smaller
\ct{BBD91} than the mass of the strange vector meson $D^*_{s}(2114)$.

In the  limits of the Borel variables and the continuum thresholds discussed
above we obtain for the total semileptonic decay width
\beq
\Gamma(D\to\kappa\ell\nu)= (5.5 \pm 1.0) \times 10^{-15} ~   
{\rm GeV} \; , 
\enq
where we have used $V_{cs}=0.97$. The same calculation done
in an -- unjustified -- zero width approximation would yield a total
semileptonic decay width which is about 20 \% larger.  

\begin{figure} \label{fig3}
\begin{center}
\epsfxsize8cm
\epsffile{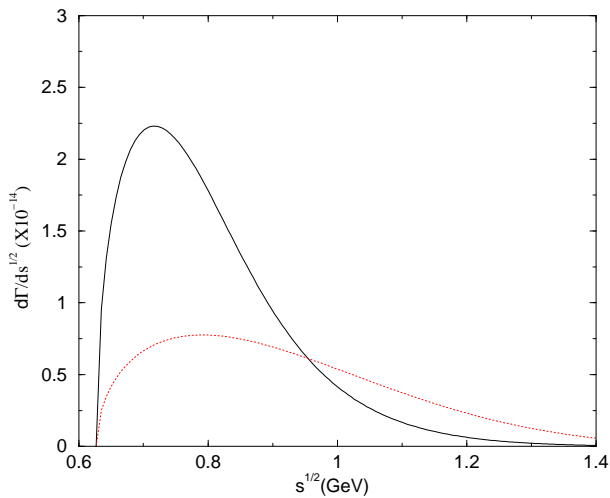}
\end{center}
\caption{Invariant mass distribution $d\Gamma(\sqrt{s})/d\sqrt{s}$ of the
hadronic final state in  the decay $D\to (K\pi)_S\ell \nu$. Solid:
resonance formation; dots: uncorrelated $K\pi$.}
\end{figure}
The spectral distribution $d\Ga/d\sqrt{s}$, Eq.~(\ref{gaspec}), where
$\sqrt{s}$ is the invariant mass of the $\pi K$ state, is given in
Fig.~3, solid line. 

Next we investigate the same decay under the assumption that the scalar
current in Eq.~(\rf{intscal}) does not couple to a resonance, but to an 
uncorrelated $\pi K$ pair in an $S$-state, {\it i.e.}, 
we use the ansatz 
(\rf{rho2}) for $\rho_\ka(s)$ entering the sum rules. In Fig.~4 
we show the dependence on the Borel variable $M_D^2$ of the decay form factor 
$f_{+\ka}(0)$ for $s_{0\kappa}= 1.6 \GeV^2$. 
Note that now the density (\rf{rho2}) describes a 
two-particle phase space and therefore the dimension of $f_{+\ka}$ is different
from the previous case. 

\begin{figure} \label{fig4}
\begin{center}
\epsfysize=8.0cm
\epsffile{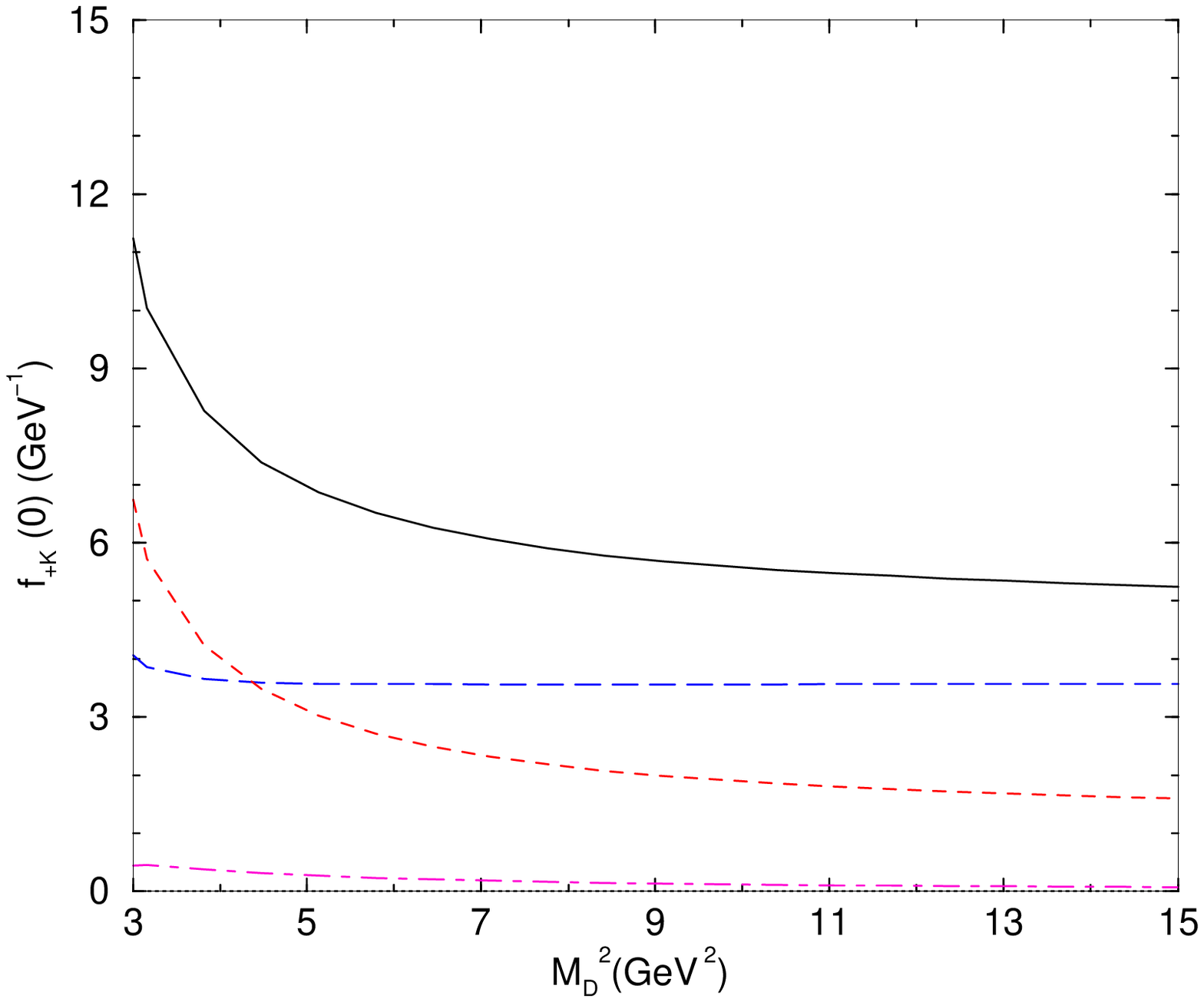}
\caption{Dependence of the form factor $f_{+\ka}$ at
$t=0$ on the Borel variable $M_D^2$. Here the decay is assumed to proceed
through an uncorrelated $\pi K$-meson pair in an $S$-state, the  density is
given by Eq.\protect (\rf{rho2}). Solid curve total contribution; long-dashed
perturbative; dashed quark condensate; dot-dashed mixed condensate; dotted
 instanton contribution. }
\end{center}
\end{figure}

The masses of the pole fit to the $t$-dependence are
practically the same as for the resonance case; for the form factor at $t=0$
we obtain
\beq
5.0 \GeV^{-1} f_{+\kappa}(0) \leq 7.1 \GeV^{-1} \; .
\enq
The total width comes out for the case of an uncorrelated $\pi K$ pair
in an $S$ state as 
\beq
\Gamma(D\to (K \pi)_S\ell\bar\nu_{\ell})=(3.7 \pm 1.1)\times
10^{-15}~{\rm GeV}.\label{deps}
\enq
The spectral distribution for this case  is also shown in
Fig.~3, with a dotted line. Although there is no resonance formation the
resulting distribution shows a maximum at approximately the mass of the
$\kappa$, which is an effect of the decrease of the total phase space 
near the kinematical limits.

The evaluation of the decay $D\to \si \ell \nu$ follows exactly the same
lines. Here the fiducial range in $M_D^2$ is chosen according to the same
criteria as before  and goes approximately from $M_D^2 = 8 \GeV^2$ to $18
\GeV^2$ corresponding approximately to a range $1 \GeV^2 \leq M_\si^2 \leq
2.3 \GeV^2$. The continuum limit $s_{0\si}$ was chosen between 1 and 1.6
$\GeV^2$. 
The resulting form factors and total decay width are, in
the case of a resonance formation with a Breit-Wigner width 
\beq
0.42\leq f_{+\si}(0) \leq 0.57   \; ,
\enq
\beq 
\Gamma(D\to \si \ell \bar\nu_{\ell}) = (8.0 \pm 2.5) \times 10^{-16} ~   
{\rm GeV} \; , 
\label{gasibw} 
\enq 
and for the case of two uncorrelated $\pi$-mesons in an $S$-state
\beq
5 \GeV^{-1} \leq f_{+\si}(0) \leq 6 \GeV^{-1}
\enq
and
\beq 
\Gamma(D\to (\pi \pi)_S  \ell \bar\nu_{\ell}) = (4.5 \pm 1.0) \times
10^{-16}  {\rm GeV}.  
\label{gasicont} 
\enq 

\begin{figure}
\leavevmode
\begin{center}
\epsfxsize8cm
\epsffile{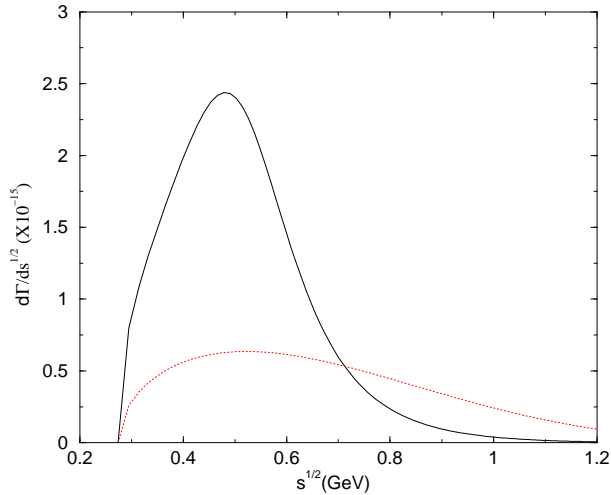}
\end{center}
\caption{The invariant mass distribution of the hadronic final state of the
rate in  the decay $D\to (\pi\pi)_S\ell \nu$. Solid: resonance formation;
dots: uncorrelated $K\pi$.}  \lb{fig5} 
\end{figure} 
The spectral distributions for both an\"satze  are shown in
Fig.~5.

\section{Summary and conclusions}
The semileptonic decays of $D$-mesons into scalar hadrons are very similar to
those into pseudoscalars. The same role played by the vector part in the
pseudoscalar case is played by the weak pseudovector for the decay into
scalars. Therefore, the theoretical expressions are very similar, only terms
proportional to the mass of the strange quark change sign. This
leads to a small reduction of the form factors compared to the decay into a
pseudoscalar state. The main difference of the semileptonic decay rates into 
$K$ and $\kappa$ is due to the different phase space. The decay into
$\sigma$ mesons is suppressed by the small value of the weak $(c,d)$  matrix
element $V_{cd}\approx 0.225$.  The spectral distributions of the invariant 
masses of the $K\pi$ and $\pi\pi$
sytems are given in Figs.~3 and 5. Their maxima are well below the
masses of the $\kappa$ and the $\sigma$, and also the forms are quite different
from Breit-Wigner distributions. The increase of the spectral ditributions is
steeper and the fall-off substantially faster than for the corresponding
Breit-Wigner forms. This is an effect of the total phase space in the
semileptonic decay. 

If the scalar current does not couple
directly to the resonance but only to an uncorrelated meson pair,
the decay rate is reduced by a factor two compared with the decay into
resonances, but nevertheless an enhancement near 0.8 and 0.5 GeV is visible due
to the total final state phase space (see Figs.~3 and 5).  With good
statistics a discrimination between the two extreme cases should be possible.
This would add valuable information on the nature of the intriguing low lying
scalar resonances. In reality things might be complicated by a coupling of the
interpolating current \rf{intscal} on the quark level to a resonance as well as
to a uncorrelated meson pair. In this case on has to construct a new density
$\rho_X$ and use it instead of the densities in Eqs.~(\rf{rho1}) or 
(\rf{rho2}), in order to
calculate the form factors $f_X(t)$.The procedure that follows is the  same.

\vspace{1cm}
 
\underline{Acknowledgements}: 
We would like to thank I. Bediaga for fruitful discussions. This work has 
been supported by CNPq, FAPERJ,  FAPESP (Brazil), and DAAD 
(Fed. Rep. Germany). 
\vspace{0.5cm}

\appendix

\section{Evaluation of the two-point sum rules}

In  Fig.~6 we display the different contributions to $A_\ka$, as a function of 
the Borel 
variable ${M'_\ka}^2$ using the ansatz in Eq.~(\ref{rho1}) for 
$\rho_\ka(s)$ and $s_{0\ka}=1.4\;\GeV^2$. As expected the instanton 
contribution is particularly important for small values of the Borel variable. 
In spite of the weak dependence of the sum rule results on ${M'_\ka}^2$, 
it is well known that radiative corrections could be large in this case. 
Therefore, we do not attach great
significance to the high stability of our results.

\begin{figure} \label{fig6}
\begin{center}
\epsfysize=8.0cm
\epsffile{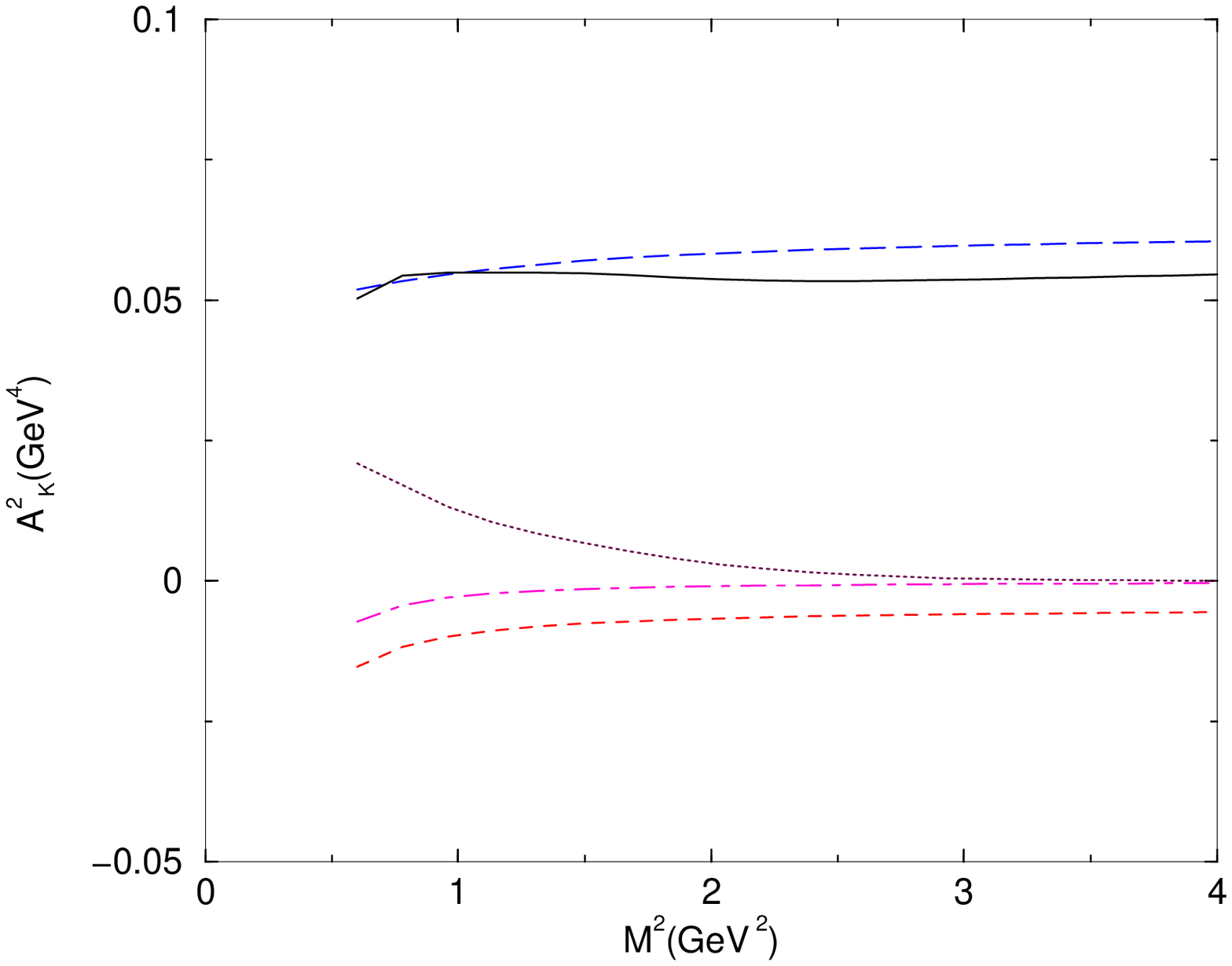}
\caption{Borel mass dependence of $A_\ka^2$. 
Solid curve: total contribution; long-dashed: perturbative; dashed:
quark condensate; dot-dashed mixed condensate; dotted : instanton
contribution.} 
\end{center}
\end{figure}

In the zero width approximation one can obtain a sum rule
for the mass by performing the logarithmic derivative of the right and
left-hand sides of Eq.~(\rf{twopointd}) with respect to $M^{-2}$, 
assuming that $f_D$ is independent of $M^{-2}$. In Fig.~7 we 
display the logarithmic derivative of the left and right-hand sides of the sum
rule (\rf{twopoint}), again assuming that $A_\ka$ is independent of the Borel
variable.  

\begin{figure} \label{fig7}
\begin{center}
\epsfysize=8.0cm
\epsffile{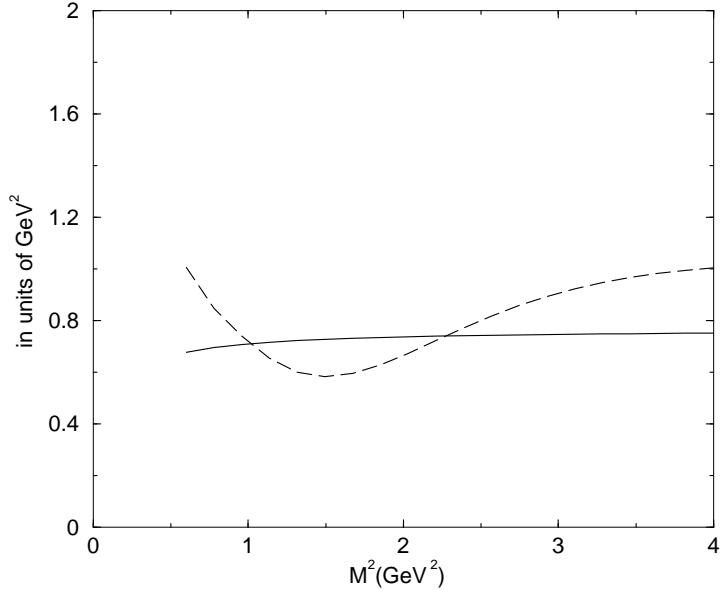}
\caption{Borel mass dependence of the rhs ( dashed line) and the
lhs ( solid line) of the sum rule generated by the logarithmic derivative of 
Eq.~(\protect\rf{twopoint}) with
respect to $M^{-2}$ for the case of the $\kappa$.}
\end{center}
\end{figure}

For
$s_{0\ka}=1.4 \GeV^2$ there is a reasonable
overlap  between the two sides of the sum rule in the Borel window $0.9\leq
M^2\leq  2 \;\GeV^2$, and this defines the  range for the Borel
variable $M^{'2}_X$ in   Eq.~(\ref{sr3}). Since all Borel variables are related
by Eqs.~(\rf{m23}) and  (\rf{mm}),  we obtain from the above range 
approximately $5 \leq M_D^2\leq 15\;\GeV^2$.

\begin{figure} \label{fig8}
\begin{center}
\epsfysize=8.0cm
\epsffile{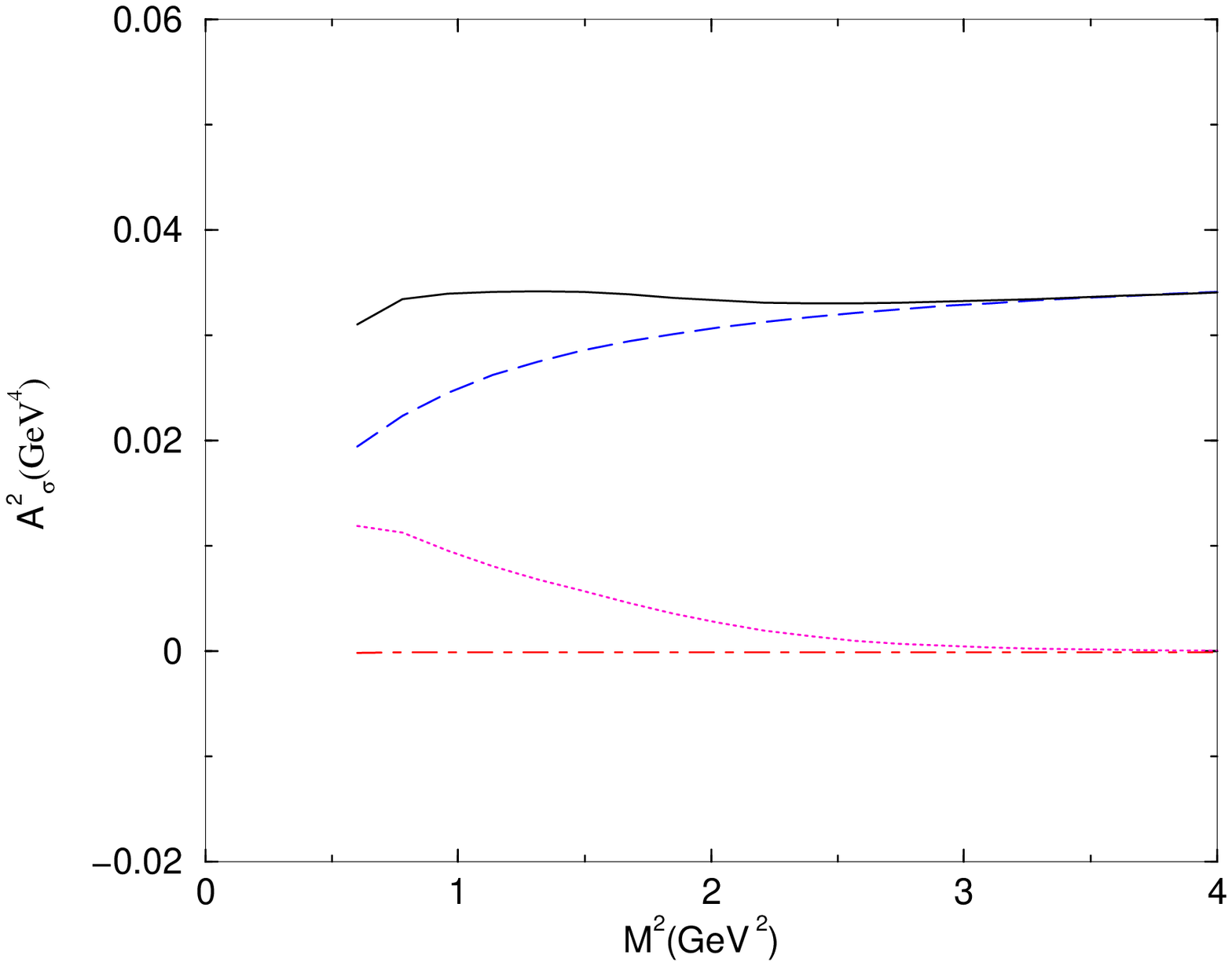}
\caption{Borel mass dependence of $A_\si^2$. 
Solid curve: total contribution; long-dashed: perturbative; dashed:
quark condensate; dot-dashed mixed condensate; dotted : instanton
contribution.} 
\end{center}
\end{figure}

In the case of $\si$, similar results are obtained and in Fig.~8 we show (for 
$s_{0\si}=1.2\,\GeV^2$ and the ansatz in Eq.~(\ref{rho1}) for $\rho_\si(s)$) 
that
the instanton contribution (dotted line) is even more important in this case,
giving a very stable result as a function of the Borel mass. The quark 
condensate and mixed condensate contributions (dashed line) are  
zero since they are now proportional to the light quark mass, which we 
take equal to zero.

In  Fig.~9 we show the lhs and the rhs of the sum rule generated by the 
logarithmic derivative of 
Eq.~(\rf{twopoint}) with
respect to $M^{-2}$, as a function of the Borel mass, using the ansatz
in Eq.~(\ref{rho1}) for $\rho_\si(s)$ and $s_{0\si}=1.2\,\GeV^2$. We see 
that there is again a good overlap
between the two sides of the sum rule in the Borel window $1.0\leq M^2\leq
2.5 \;\GeV^2$. Since the mass of the particle is related with the square root
of the lhs of this sum rule, from this figure we see that
our result is  compatible with $m_\si=0.5\, \GeV$ found in \cite{Ait01}.

\begin{figure} \label{fig9}
\begin{center}
\epsfysize=8.0cm
\epsffile{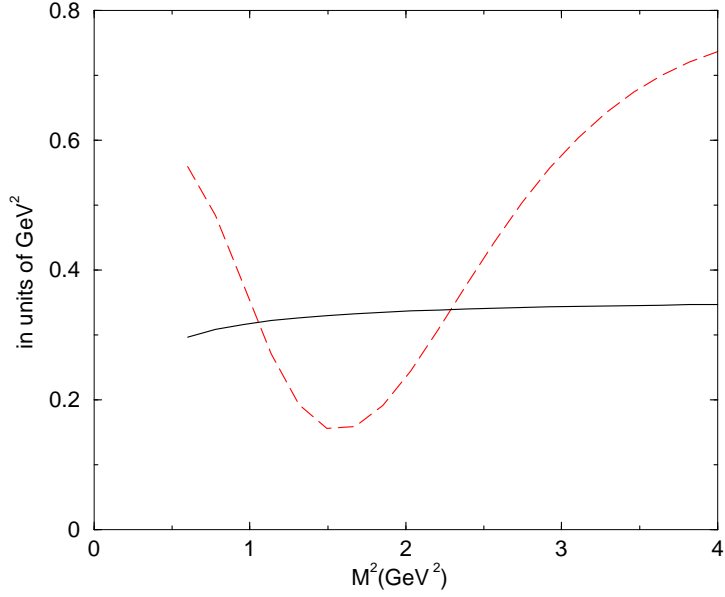}
\caption{Borel mass dependence of the rhs ( dashed line) and the
lhs ( solid line) of the sum rule generated by the logarithmic derivative of 
Eq.~(\protect\rf{twopoint}) with
respect to $M^{-2}$ for the case of the $\sigma$.}
\end{center}
\end{figure}

\section{Perturbative and nonperturbative Contributions to the two-
and three-point functions}

In all this work we take into account the mass of the strange quark at most 
squared and neglect the mass of the light quarks. For the scalar meson $X$,
we consider the particular case of $\kappa$, since the $\sigma$ can be easily
obtained from it by neglecting the strange quark mass. The perturbative 
contributions for the two-point functions defined 
in Sec.~III are:
\beq
\sigma_D(s)={3\over8\pi^2}{(s-m_c^2)^2\over s}\; 
\enq
and 
\beq
\sigma_\ka(s)={3\over8\pi^2}(s-2m_s^2)\;.
\enq

The nonperturbative contributions including the quark and mixed 
condensates are
\beq
\hat K_D(M^2)=-m_c\lag\bar{q}q\rag e^{-m_c^2/M^2}\left[1+{m_0^2\over 2M^2}
\left(
1-{m_c^2\over2M^2}\right)\right]\;
\enq
and
\beq
\hat K_\ka(M^2)=m_s e^{-m_s^2/M^2}\left(\lag\bar{q}q\rag+{ \lag\bar{s}s\rag
\over2}+{m_0^2\over 2M^2}\lag\bar{q}q\rag\right)\;,
\enq
where we have defined $\lag\bar{q}g_s\sigma.Gq\rag=m_0^2\lag\bar{q}q\rag$. The
instanton induced contribution is given by
\beq
\hat K_I(M^2)={\bar{n}\over2\bar{m}_s\bar{m}_q}M^2z^2\int_{z^2/4}^\infty\;
dx{x^2\over(x-z^2/4)^2}e^{-x^2\over x-z^2/4}\;,\label{inst}
\enq
where $z=M\bar{\rho}$, and we have introduced the average instanton 
size, $\bar{\rho}$, and the instanton number density, $\bar{n}$, given by
\cite{shu}
\begin{equation}
\bar{n} 
\simeq \frac12 {\rm fm}^{-4}, \quad \quad \bar{\rho} \simeq \frac13 
{\rm fm}. \label{n}
\end{equation}
In Eq.~(\ref{inst}) $\bar{m}_s$ and $\bar{m}_q$ are the effective quark
masses which we take to be $\bar{m}_s=400\, \MeV$ and  $\bar{m}_q=300\, \MeV$.

The perturbative double spectral function is obtained by using the Cutkosky
rules and, in the case of the meson $\kappa$, is given by
\beqa
\rho_{+\kappa}(s_D,s_X,t)&=&{-3\over8\pi^2\la^{3/2}}\left[-2s_Ds_Xt+m_c^2s_X(
s_D-s_X+t)+m_s^2s_D(s_X-s_D+t)+m_cm_s\right.\nn\\
&\times&\left.\left(m_c^2(s_D-s_X-t)+m_s^2(s_X-s_D-t)
-(s_D-s_X)^2+t(s_D+s_X)\right)\right]\nn\\
&\times&\Theta(s_D-m_c^2)\Theta(s_X-s_{Xmin})\Theta(s_{Xmax}-s_X)\;,
\lb{per3}
\enqa
where
\beq
s_{Xmin}={m_c^2(-m_c^2+m_s^2+s_D+t)+m_s^2s_D-s_Dt+(m_c^2-s_D)
\sqrt{\la(m_c^2,m_s^2,t)}\over2m_c^2}\; 
\enq
and
\beq
s_{Xmax}={m_c^2(-m_c^2+m_s^2+s_D+t)+m_s^2s_D-s_Dt-(m_c^2-s_D)
\sqrt{\la(m_c^2,m_s^2,t)}\over2m_c^2}\;.
\enq

The nonperturbative contributions, including the quark and mixed 
condensates, which survive the double Borel transformation in $p_D^2$
and $p_X^2$ are
\beqa
\hat K_{+\kappa}(M_D^2,M_X^2,t)&= &\lag\bar{q}q\rag e^{-m_c^2/M_D^2}
\left[{m_c-m_s\over2}-m_0^2\left(m_c^2{m_c-m_s\over8M_D^4}
\right.\right.\nn\\
&-&{2 m_c-m_s\over6 M_D^2}+{(4 m_c^2+m_c m_s-2 t)(m_c-m_s)\over24M_D^2 M_X^2}
\nn\\
&-&\left.\left.
{m_c-2 m_s\over6 M_D^2}+{(m_c m_s-2 t)(m_c-m_s)\over24M_D^2 M_X^2}\right)
\right]\;,
\enqa
and the instanton induced contribution is given by
\beqa
\hat K_{+I}(M_D^2,M_X^2,t)&= &{8{\bar\rho}^4t\over\pi^2}{\bar{n}\over 
\bar{m}_s\bar{m}_q}
\int_0^\infty dr~ ds~ du~ d\ga~
{(\ga M_D^2-1)M_X^4\over(1+4\ga s)^2(\ga s M_D^2-4M_X^2)^2\ga s^2}\nn\\
&\times&{\sqrt{r u\over(s-r)(A_0-u)}}~e^{-{\bar\rh}^2(r+u)}~e^{-\ga m_c^2}~
e^{-\al_0{\bar\rh}^2}~e^{-tA_0\over sM_D^2}\;,
\enqa
where
\beq
A_0={s(\ga M_D^2-1)\over 1+4s\ga}\;
\enq
and
\beq
\al_0={s+(1+4s\ga)t)M_X^2\over s+(1+4s\ga)t)-4M_X^2}\;.
\enq


\begin{thebibliography}{99}
\bibitem{PDG00} D.E. Groom et al. (Particle Data Group),{\em  Eur. Phys. Jour.}
C15:437, 2000.
\bibitem{Ait01} E.M. Aitala et al. {\em Phys. Rev. Lett.} 86:770, 2001
\bibitem{e791} C. G\"obel, (for the E791 Collaboration), hep-ex/0110052.
\bibitem{GNPT00} R. Gatto, G. Nardulli, A.D. Polosa and N.A. Tornqvist,
{\em Phys.\ Lett.}, B494:168, 2000
\bibitem{Lin02} J.M. Link et al (FOCUS collaboration), hep-ex/0203031
\bibitem{SVZ79} M.A. Shifman, A.I. and Vainshtein and V.I. Zakharov,
{\em Nucl. Phys.}, B147:385, 1979

\bibitem{CK00} P. Colangelo and A. Khodjamirian,
     {\em QCD sum rules: A modern perspective}, in {\em At the Frontier of
Particle Physics}, ed. M. Shifman, Singapore 2001, hep-ph/0010175;

\bibitem{JOP00}
M. Jamin, J~A. Oller, and A. Pich.
\newblock {\em Nucl. Phys.}, B587:331, 2000.
\newblock {\em Nucl. Phys.}, B622:279, 2002.

\bibitem{BBD91} P. Ball, V.M. Braun and H.G. Dosch, {\em Phys. Rev.},
D44:3567, 1991.


\bibitem{shu} E.~V. Shuryak, Nucl.\ Phys.\ 
{\bf B203}, 93, 116, 140  (1982); T. Schaefer and E. Shuryak, {\em Rev. Mod.
Phys.}, 70:323, 1998.

\bibitem{bro} D.J. Broadhurst, {\em Phys. Lett.}  B101:423, 1981.

\bibitem{BBB98}
E.~Bagan, Patricia Ball, and V.~M. Braun.
\newblock {\em Phys. Lett.}, B417:154, 1998.

\bibitem{BBFG95}
E.~Bagan, Patricia Ball, B.~Fiol, and P.~Gosdzinsky.
\newblock {\em Phys. Lett.}, B351:546, 1995.
\end{thebibliography}
\end{document}